\documentclass[%
aps,prx,
amsmath,amssymb,
reprint,
superscriptaddress]{revtex4-1}

\usepackage[utf8]{inputenc}
\usepackage[T1]{fontenc}
\usepackage{mathptmx}
\usepackage{textcomp}

\usepackage{graphicx}
\usepackage{hyperref}
\usepackage{color}
\usepackage{ulem}

\begin{document}
\newcommand{\Jc}{$J_{c}$}
\newcommand{\Ic}{$I_{c}$}
\newcommand{\YGBCO}{Y$_{0.77}$Gd$_{0.23}$Ba$_2$Cu$_3$O$_7$}
\newcommand{\BHO}{BaHfO$_3$}
\newcommand{\Jcsf}{$J_{c}^{s.f.}$}
\newcommand{\mum}{\textmu{}m}
\newcommand{\muV}{\textmu{}V}
\newcommand{\mus}{\textmu{}s}

\title{Novel dynamics and critical currents in fast superconducting vortices at high pulsed magnetic fields}

\author{Maxime Leroux}
\affiliation{Materials Physics and Applications Division, Los Alamos National Laboratory, Los Alamos, NM, USA}

\author{Fedor F. Balakirev}
\affiliation{Materials Physics and Applications Division, Los Alamos National Laboratory, Los Alamos, NM, USA}

\author{Masashi Miura}
\affiliation{Graduate School of Science and Technology, Seikei University, Tokyo, Japan}

\author{Kouki Agatsuma}
\affiliation{Graduate School of Science and Technology, Seikei University, Tokyo, Japan}

\author{Leonardo Civale}
\affiliation{Materials Physics and Applications Division, Los Alamos National Laboratory, Los Alamos, NM, USA}

\author{Boris Maiorov}
\affiliation{Materials Physics and Applications Division, Los Alamos National Laboratory, Los Alamos, NM, USA}

\date{\today}

\begin{abstract}
Non-linear electrical transport studies at high-pulsed magnetic fields, above the range accessible by DC magnets, are of direct fundamental relevance to the physics of superconductors, domain-wall, charge-density waves, and topological semi-metal. All-superconducting very-high field magnets also make it technologically relevant to study vortex matter in this regime. 
However, pulsed magnetic fields reaching 100\,T in milliseconds impose technical and fundamental challenges that have prevented the realization of these studies. 
Here, we present a technique for sub-microsecond, smart, current-voltage measurements, which enables determining the superconducting critical current in pulsed magnetic fields, beyond the reach of any DC magnet. 
We demonstrate the excellent agreement of this technique with low DC field measurements on \YGBCO\ coated conductors with and without \BHO\ nanoparticles.
Exploring the uncharted high magnetic field region, we discover a characteristic influence of the magnetic field rate of change ($dH/dt$) on the current-voltage curves in a superconductor. We fully capture this unexplored vortex physics through a theoretical model based on the asymmetry of the vortex velocity profile produced by the applied current.

\end{abstract}


\maketitle
\section{Introduction}
Non-linear current-voltage (I-V) curves at high fields are key to study the breaking of Ohm's law in topological (semi)-metals\cite{Ramshaw,Shin2017}, to help identify the hidden order nature in URh$_2$Si$_2$,\cite{IVURu2Si2} and to explore the dynamics of charge density waves at very high magnetic fields above the 3D-2D transition\cite{Gerber3DCDWYBCOhighB}.
Non-linear I-V studies also bring essential information about the dimensionality, disorder interaction and nature of the superconducting vortex solid as well as the critical phenomena associated with vortex solid-liquid phase transition \cite{BlatterRMP,nelsonnew,boseh}. 

High fields studies are essential to understand high temperature superconductors, because of their immense characteristic fields (>100\,T) that cannot be reached using conventional DC magnets.
The advent of hydrogen-based superconductors \cite{HSuperPRL2019,DrozdovHSup,Fedor2019Hsup} with critical temperatures in excess of $260~$K heightens the already present need created by Cu- and Fe-based superconductors. 
Depending on the superconductor properties (electronic mass anisotropy, $T_c$, etc) and the type and density of material disorder,
the vortex solid phase changes drastically from crystalline to diverse glass phases \cite{BlatterRMP,nelsonnew,scottsmectic}
or even to emergent novel phases like Fulde–Ferrell–Larkin–Ovchinnikov.\cite{FFLO_Ikeda} 

The study of vortex pinning in commercially relevant superconductors at high fields is key for developing magnets and power applications such as the recent 32\,T record in an all-superconducting magnet\cite{ObradorsPuigReview,FoltynNatMat,progress32Tinsert,32Tmaglosses}. 
Vortex pinning arises from the presence of defects in the superconducting material that immobilize vortices, allowing electric currents to flow without moving vortices (i.e. without dissipation) as long as the applied current density is less than a critical value, called critical current density (\Jc). The value of \Jc\ is set by the interplay between vortex-vortex and vortex-defects interactions\cite{BlatterRMP,CampbellEvetts2001,borisnatmat,civalejltp}.  
Although REBa$_2$Cu$_3$O$_7$ high-temperature superconductors (REBCO, where RE is a rare earth element) are the materials with the highest \Jc, most knowledge on vortex pinning is reduced to around 20\% of REBCO's magnetic field - temperature phase diagrams\cite{FoltynNatMat,ObradorsPuigReview}. The same limitation will apply for H-based superconductors as studies now focus on critical fields properties.\cite{HSuperPRL2019,DrozdovHSup,Fedor2019Hsup}  This limitation makes vortex pinning in superconductors at fields above 20\,T almost uncharted territory\cite{Larbalestier35Tinsert, Xu31TAPLMat2014,Larbalestier2014,Abraimov2015,Selva2017Je31Tesla}. 
So, a sensible pathway forward is to achieve \Jc\ measurements in pulsed magnetic fields\cite{Bi2212Jcpulsedfield, wozniak2009comparison, wozniak2012characterisation}, as peak fields up to 100\,T are routinely achievable in the millisecond range at large facilities\cite{nhmfl,100TToulouse,BATTESTI2018} and 30\,T table-top systems are available.\cite{30Tpulse} 
 
Several technical and fundamental obstacles need to be cleared first. The measurement of \Jc\ in a superconductor typically involves measuring an I-V curve i.e. applying an electric current while monitoring that the voltage across the sample does not exceeds a set threshold.
For practical reasons, in commercial superconductors $J_c$ is defined via an electric field criterion $E_c$ in the I-V curve. Note that this definition of $J_c$ is different from the fundamental definition of $J_c$ which corresponds to the limit for the depinning of vortices.
Previous attempts to measure \Jc\ in pulsed magnetic fields involved applying a single current pulse per magnetic field pulse\cite{NbSnpulsedcurrent,Bi2212Jcpulsedfield, wozniak2009comparison, wozniak2012characterisation, MGOpulsedcurrent}.  
This method does not produce an I-V curve, but rather a critical current defined as the onset of an abrupt (potentially destructive) increase in the resistivity of the sample.  
This requires several magnet pulses to produce a single \Jc\ data point with a high risk of destroying the sample. Thus a major technical roadblock to be solved is measuring I-V curves with sufficient resolution in time (sub-\mus) and voltage (\muV), coupled with on-the-fly monitoring and decision making to avoid destroying the sample by passing too much current.

Previous studies also did not fully address a fundamental problem associated with the physical interpretation of the results. 
During the period that the current is applied (typically milliseconds), the magnetic field amplitude can change significantly (>1\,T) and at very large rates ($\mu_0 dH/dt\approx 1000$\,T/s). The large $dH/dt$ is particularly troublesome, because vortices are rapidly entering and then exiting the sample while the voltage is recorded. Thus, it is unclear whether the standard situation in I-V curve measurements in DC fields, i.e. that vortices are motionless until the applied current unpin them, occurs in this case.

Here we show that it is possible to measure reproducible I-V curves, in a 65\,T pulsed magnet, using Fast Programmable Gate Arrays (FPGA) with 50\,\textmu{}m wide superconducting thin films on metal substrates. This enables us to reliably determine \Jc\ at the largest field ever (50\,T) in a superconductor. The \Jc\ values measured compare quantitatively well with the results obtained at low DC fields in the same samples.
In addition, we find that the rate of change of the magnetic field can strongly affect the I-V curves. We present a theoretical model that quantitatively reproduces this response by taking into account the asymmetry of vortex motion when applying a DC current to a superconductor in a time-dependent magnetic field.
Our study thus lays the foundations for widespread adoption of critical current measurements in pulsed fields, while also unlocking the experimental access to new high-field, high-rate, vortex physics.

\section{Results}
\subsection{\Jc\ measurements in pulsed fields}
\begin{figure}
\includegraphics[width=\columnwidth]{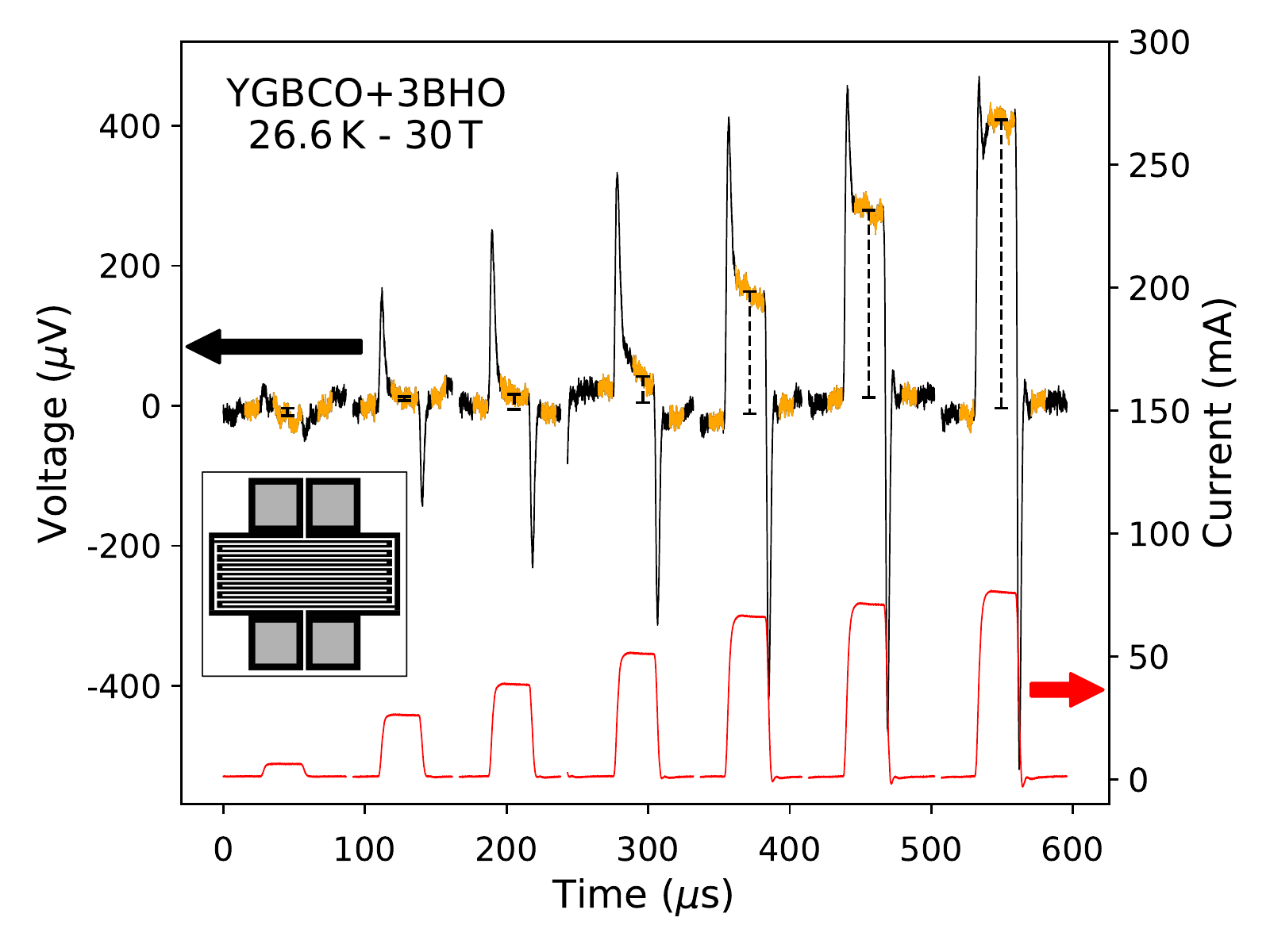}%
\caption{\label{fig:raw_pulses}\textbf{FPGA based, fast I-V curves measurements.}
Source current pulses (red lines) and corresponding raw voltage pulses (black lines) in
YGBCO+3BHO at 26.6\,K, taken at the peak field of 30\,T magnetic field pulses.
Each pulse of current is $\approx 25$\,\mus\ long.
The regions highlighted in orange on the voltage pulses indicate the regions over which the voltage is averaged and over which the background is subtracted.
The black dashed lines represent the resulting average amplitude of the voltage above the background.
(Inset) schematic of the sample meandering pattern.
}
\end{figure}  

\begin{figure}
\includegraphics[width=\columnwidth]{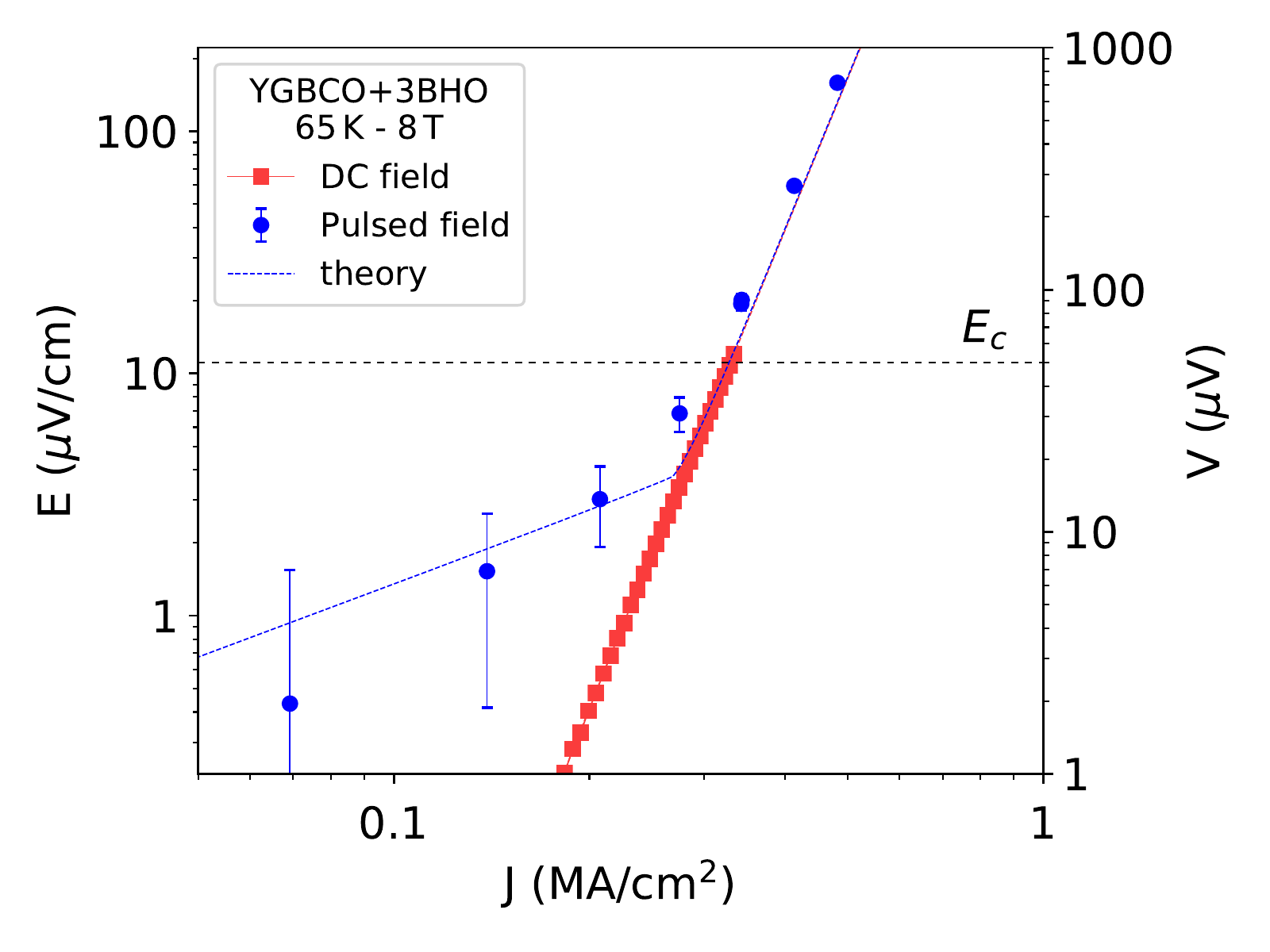}
\caption{\label{fig:DCvsPulsed}\textbf{Agreement between DC and pulsed field $\mathbf{J_\mathrm{c}}$.} I-V curves measured at 65\,K and 8\,T in DC field and 64.6\,K and 8\,T in pulsed field, in the sample with BHO nanoparticles.
Both curves show a power-law dependence with similar $n$-value and both curves yield very close values of the critical current using a 11\,\muV/cm criterion.
The deviation below $\approx10$\,\muV\ is due to $dH/dt$ effects in pulsed fields: the dashed blue line is the theoretical curve for $\mu_0 dH/dt=-15$\,T/s.
}
\end{figure}

We measured a standard \YGBCO\ (YGBCO) sample as well as a \YGBCO\ sample with 3\% by weight of \BHO\ nanoparticles (YGBCO+3BHO, see Methods section).
Fig.~\ref{fig:raw_pulses}\ shows the time dependence and simultaneous measurements of a typical series of voltage and current pulses on the YGBCO+3BHO sample. The noise floor on the raw signal is $\sim 20$\,\muV\ at the acquisition rate of 125\,MHz, i.e. 1.8\,nV$/\sqrt{\mathrm{Hz}}$, close to the input noise of the preamplifier
(see Methods).
The positive (negative) voltage spikes at the front (end) of each pulse are due to the inductive response of the sample strip to the ramping current.
After the field pulse, the initial voltage measurements (embedded in the FPGA for sample protection, see Methods) is refined by carefully reanalyzing each current pulse averaging and background subtraction procedures.
Using the voltage determined for each current pulse (black vertical dashed lines), we can then plot an I-V curve, from which \Jc\ is determined using a criterion $E_c=11$\,\muV/cm (i.e. $50$\,\muV, see Fig.\ref{fig:DCvsPulsed}).
The I-V curves thus measured are very reproducible from pulse to pulse, and are also robust against variations of experiment parameters such as the voltage integration time, and the size and order of the steps of current (see Methods). This enables to reconstruct a complete I-V curve from data taken in different magnetic field pulses.

\begin{figure}
\includegraphics[width=\columnwidth]{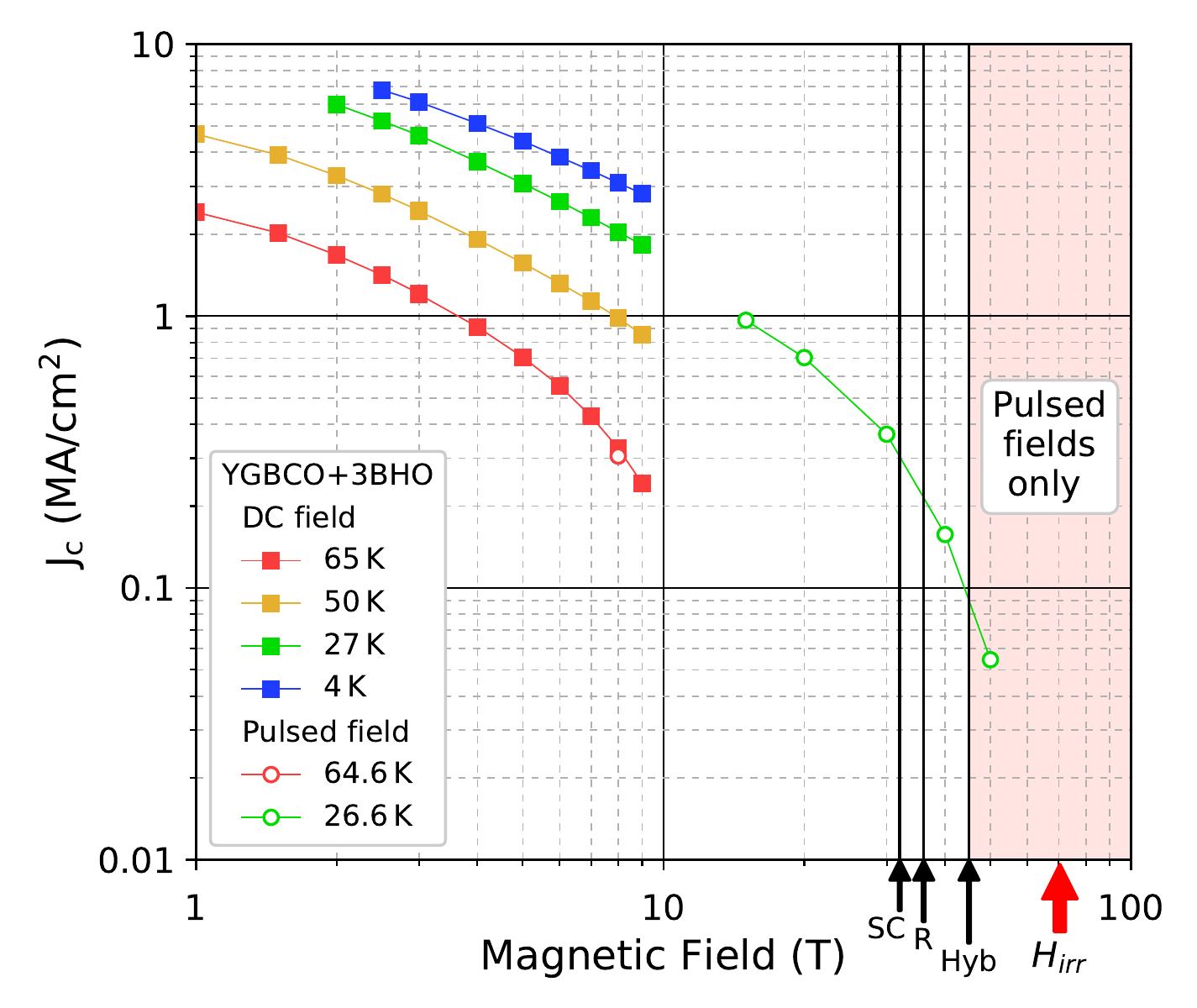}
\caption{\label{fig:JcH}\textbf{Magnetic field dependence of the critical current up to 50\,T.} Critical current of the sample with BHO nanoparticles, using a 11\,\muV/cm criterion (50\,\muV), measured in DC field up to 9\,T in a PPMS system and in pulsed field up to 50\,T at the NHMFL pulsed field facility.
Data at 65\,K and 8\,T shows a good quantitative agreement between DC and pulsed field measurements. Data at high field is also in-line with DC field results. $H_{irr}$ is determined by the field at which $J_c\approx7\times 10^{-5}$\,MA/cm$^2$.
 Labels indicate typical maximum field values achieved with current magnet technologies (SC: all-superconducting, R: resistive, Hyb: hybrid). At present, only pulsed field measurements are possible for field values above 45\,T.}
\end{figure}

\begin{figure}[t]
\includegraphics[width=\columnwidth]{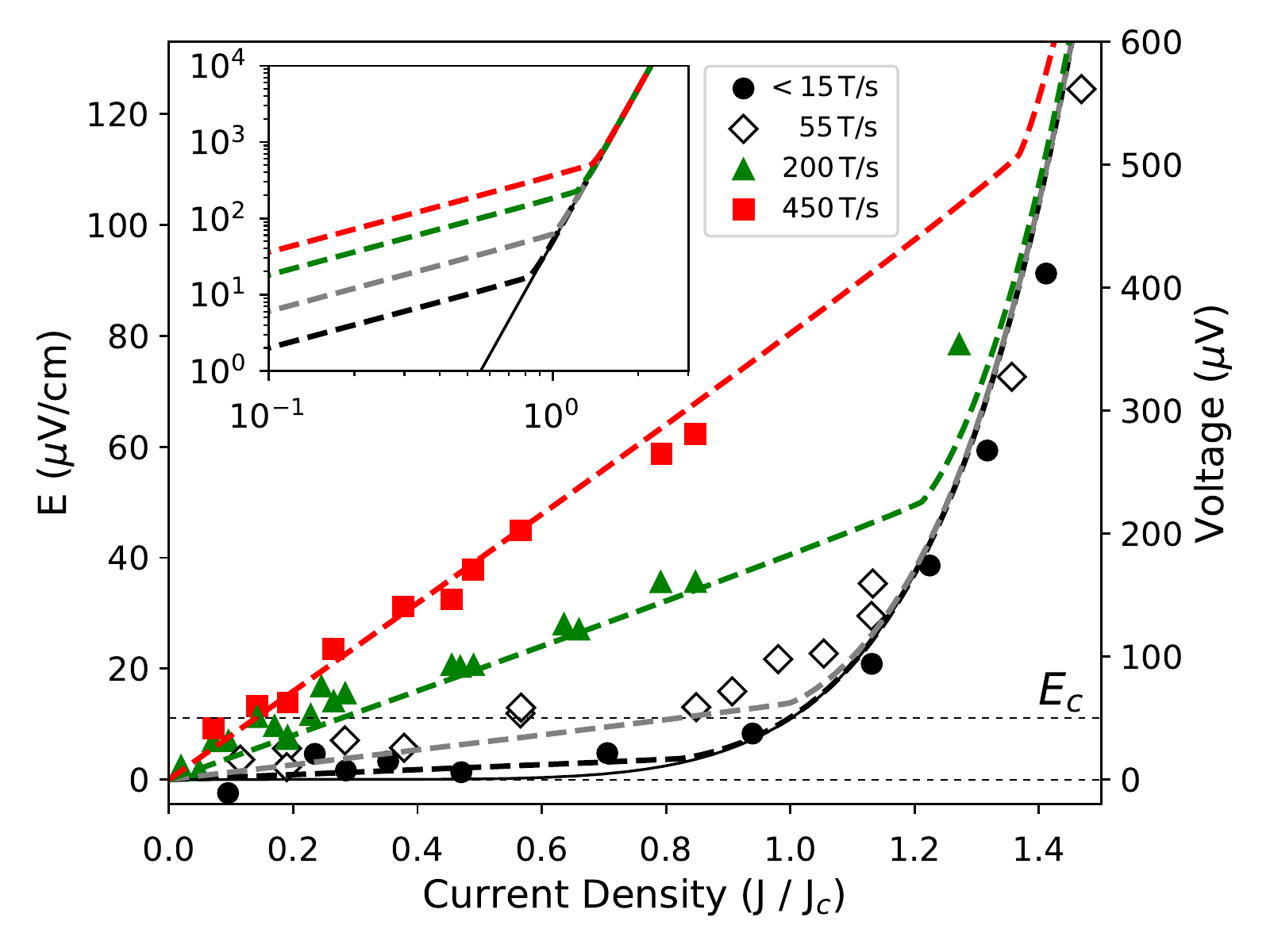}
 \caption{\label{fig:IVBdot}\textbf{Dependence of the I-V curves on the magnetic field rate of change $\mathbf{\dot H \equiv dH/dt}$.}
 \textbf{(symbols)} Experimental points in sample YGBCO+3BHO at 26.6\,K and 30\,T peak field, which show a clear dependence on $\dot{H}$. For large $\dot{H}$ values the I-V curve is initially ohmic (linear), whereas data at low $\dot{H}$ ($\lesssim 15 $\,T/s, closest to peak field) show the expected power-law I-V curve from which the critical current can be measured. 
\textbf{(dashed lines)} A theory of vortex motion from the interplay of $\dot{H}$ and a DC current (see Eq.~\ref{Icalc}) can quantitatively reproduce this behavior. \textbf{(solid black line)} Theoretical curve for $\dot{H} = 0$.
\textbf{(inset)} Log-log plot of the theoretical curves shown in the main axes. }
\end{figure}
 
In Fig.~\ref{fig:DCvsPulsed} we compare I-V curves at 8\,T for YGBCO+3BHO measured in DC field at 65\,K and in pulsed fields at 64.6\,K. For $V>10$\,\muV, these results evidence the very good agreement on the determination of \Jc, as well as on the slope ($V\propto I^n$,  $n$-value), between DC and pulsed field data. The agreement is similarly good in the YGBCO sample.
However, for $V<10$\,\muV, there is a small but clear systematic deviation of the pulsed field data.
We will see in the next section that this deviation relates to $dH/dt$ effects and that it gets larger as $dH/dt$ increases. So, for the rest of this section we focus on I-V curves measured with $\mu_0 dH/dt\leq15$\,T/s that do not affect the determination of $J_c$.
 
We now examine the magnetic field range where I-V curves have never been measured before in a superconductor. At $T=26.6$\,K we were able to measure I-V curves up to 50\,T (Fig.~\ref{fig:JcH}) and extract $J_c$. The \Jc$(H)$ data taken in pulsed fields follows the field dependence observed in DC fields measured up to 9\,T, with a similar $\alpha$ value ($J_c \propto H^{-\alpha}$). Above 30\,T, a more pronounced decrease in $J_c(H)$ is observed, similarly to what is found in REBCO tapes at higher temperatures near the irreversibility field\cite{civaleapl,borisnatmat,civalejltp}. The rapid decrease in $J_c(H)$ points out the importance of increasing the irreversibility field $H_{irr}$ in this range of temperatures, as it can drastically affect \Jc\ in the 20\,--\,40\,T range\cite{borisba122,miura}. This is the first $J_c$ determination at magnetic fields above those achievable using DC magnets, thus opening up a new frontier in vortex matter research.

\subsection{Vortex physics in a large dH/dt}

Determining \Jc\ from an I-V curve assumes that vortices are pinned and then moved by the force produced by the applied DC current. Stated otherwise, it assumes that over the timescale of the I-V measurement, vortex motion is caused predominantly by the DC current and not as a result of the fast changing magnetic field.
In this section, we analyze what is the maximum $dH/dt$ ($\equiv \dot{H}$) that still fulfills this condition, and how the I-V curves are affected above that threshold.

Experimentally, the effect of a large $\dot{H}$ on the I-V curves is quite striking.
At 30\,T, 26.6\,K for YGBCO+3BHO (Fig.~\ref{fig:IVBdot}), we find that at low $\dot{H}$ the I-V curves shape is the expected power-law with a well defined \Jc, whereas at higher $\dot{H}$ the initial part of the I-V curve becomes ohmic (linear in $J$). The power-law behavior is recovered at high $J$, as can be seen for the $\dot{H} =55$ and $200$\,T/s I-V curves and inferred for the curve at 450\,T/s.  In terms of vortex velocity, the ohmic behavior at $\dot{H}= 450$\,T/s in Fig.~\ref{fig:IVBdot}, corresponds to a resistivity of $\sim 0.19$\,n$\Omega.$cm, which is 5 orders of magnitude smaller than the Bardeen-Stephen flux-flow resistivity limit $\rho_{ff} = \rho_n * (H/H_{c2}) =  16.2$\,\textmu{}$\Omega$.cm, thus indicating a much slower movement of vortices (see Methods for details).  We find the same behavior on the I-V curves of the YGBCO sample at 50\,K, 10\,T. 
The theoretical model that we present hereafter is able to completely capture the changes with $\dot{H}$, using only the $J_c$ and $n$-value obtained at low $\dot{H}$ and the geometry of the sample, as shown by dashed lines in Fig.~\ref{fig:IVBdot}.

We model vortex motion in the presence of both a DC current and a large $\dot{H}$.
We consider a long superconducting slab of width $W$ along $\vec{e_x}$, thickness $d$ along $\vec{e_z}$ and length $L$ along $\vec{e_y} $ (such that $-W/2 < x < W/2$ and $L\gg W \gg d$) with \textbf{H}$\parallel \vec{e_z}$.
We assume a Bean model\cite{BeanPRL1962} ($J_c$ independent of $B$, over the range of $B$ values present in the sample), the usual constitutive equation $E = E_c \cdot (J/J_c)^n$, and focus on the fully penetrated critical state before and after the maximum magnetic field of the pulse. Demagnetizing effects are negligible at the high fields of our study. In the ideal case (without flux creep, $n=\infty$), $B(x)$ and $J(x)$ would have the standard linear and step-like profiles across the width $W$ (Ref.~\cite{tinkham1980introduction}, p.174).

We now calculate $B(x)$ and $J(x)$ for finite $n$.
The number of vortices in the sample changes only because of vortices entering or exiting through the surface. Thus, when $H$ decreases by $\delta H\ll H$ (resp. increases), the vortices at $0 \leq u \leq x$ expand the region they occupy to $0 \leq u \leq x+\delta x$ (resp. contract to $0 \leq u \leq x-\delta x$). From this, the flow of vortices induced by $\dot{H}$ in the bulk can be described by a standard flow-conservation equation:
$
\mu_0 \dot{H} + \frac{\partial \left( B\, v \right)}{\partial x} = 0
$
\!. Then as $v(0)=0$ by symmetry (the vortices at the center of the slab do not move), the vortex velocity induced by $\dot{H}$ is:
$$
\vec{v}(x) = -\frac{\mu_0 \dot{H}}{B(x)}\,x \, \vec{e_x}
$$
which show the vortices near the edges move the fastest, in opposite directions from both sides (e.g. $\dot{H}<0$ implies $v>0$ for $x>0$ as the vortices flow out).
As each moving vortex carries a flux $\phi_0$, in the frame of reference where the superconductor is at rest the vortices induce an electric field:
$
\vec{E}_{\dot{H}}  = - \vec{v} \times \vec{B} = - \mu_0 \dot{H} \, x \, \vec{e_y}
$
yielding the current density profile:
$
\vec{J}_{\dot{H}}(x) = J_c \, \left| \frac{\mu_0 \dot{H}x }{E_c} \right|^{1/n} \mathrm{sign}(-\dot{H}x ) \, \vec{e_y}
$
\!, and the magnetic field profile from Amp\`ere-Maxwell equation $\partial B_z/\partial x = -\mu_0 J_y $ :
$$
\vec{B}(x) = \mu_0\vec{H} - \mu_0 J_c \frac{n}{n+1} \left| \frac{\mu_0 \dot{H}}{E_c}\right|^{\frac{1}{n}} \left[ \left(\frac{W}{2}\right)^{\frac{n+1}{n}} - \left|x\right|^{\frac{n+1}{n}} \right]\vec{e_z}
$$
For $n\rightarrow \infty$ the usual linear $B(x)$ and step-like $J(x)$ profiles are found, with no dependence on $\dot{H}$. In the absence of applied current, because of the symmetry around $x=0$, the electric field produced by the exit (or entrance) of vortices does not generate a net $E$ along the length $L$ of the superconductor as the contributions from each side cancel out.

\textbf{However, when a DC current $I$ is applied, this symmetry is broken.}
Here, we assume that the total electric field is the superposition of that induced by $\dot{H}$ and the uniform ``$E_J$'' induced by the DC current when no $\dot{H}$ is present:
$
\vec{E} = \vec{E}_J + \vec{E}_{\dot{H}} = \left( E_J - \mu_0 \dot{H} x \right)\, \vec{e_y}
$
\!. As a consequence, the voltage measured along L is still that originating from $E_J$:
$
V = \int^L_0 1/W \int^{W/2}_{-W/2} \left( E_J - \mu_0 \dot{H} x \right)\, \mathrm{d}x \mathrm{d}y = E_J L
$
but the current density profile is fundamentally modified because of the non-linear $E$-$J$ relation:
$$
\vec{J}_{\dot{H}}(x) = J_c \, \left| \frac{E_J - \mu_0 \dot{H}x }{E_c} \right|^{1/n} \mathrm{sign}(E_J -\mu_0\dot{H}x ) \, \vec{e_y}
$$
which by integration, yields an analytical expression for the I-V curves: 

\begin{equation}
I = \frac{d J_c}{\mu_0 \dot{H} E_c^{\frac{1}{n}}} \frac{n}{n+1} \left[ \left| \frac{V}{L} + \mu_0 \dot{H} \frac{W}{2} \right|^{\frac{n+1}{n}} - \left| \frac{V}{L} - \mu_0 \dot{H} \frac{W}{2} \right|^{\frac{n+1}{n}} \right]
\label{Icalc}
\end{equation}
This expression reveals the electric field scale $\epsilon = \frac{1}{2}W\mu_0 |\dot{H}|$.
 
For $E_J\gg \epsilon$ we recover the standard power-law $E/E_c = (J/J_c)^n$, whereas for $E_J \ll \epsilon $ we find the initial, $\dot{H}$-dependent, linear I-V curve observed experimentally.
Expanding Eq.~\ref{Icalc}\ \ for $ E_J \ll \epsilon $, we find the theoretical effective resistance of the initial linear behavior:
\begin{equation}
R_{eff} = \frac{L}{W^{\frac{1}{n}} d } \frac{E_c^{\frac{1}{n}}}{J_c} \left| \frac{\mu_0 \dot{H}}{2} \right|^{\frac{n-1}{n}}
\label{Reff}
\end{equation}
which goes as $\left| \dot{H} \right|^{\frac{n-1}{n}}$, i.e. almost proportional to $\dot{H}$ for  $n\gg 1$.
Dashed lines in Fig.~\ref{fig:IVBdot} display several of I-V curves calculated from Eq.~\ref{Icalc} using the values: $d=290$\,nm, $ W = 50$\,\mum, $L= 4.51$\,cm, $E_c =11$\,\muV/cm, $J_c=0.37$\,MA/cm$^2$, $n= 6.67$, and $\mu_0 \dot{H} = 0,\ 15,\ 55,\ 200,\ 450$\,T/s.
These theoretical curves show a remarkable quantitative agreement with the experimental data. We emphasize that there are NO free parameters for these curves as $J_c$ and $n$
were determined from the high-$E$/high-$J$ power-law behavior for $|\dot{H}|<15$T/s.

\begin{figure}
\includegraphics[width=\columnwidth]{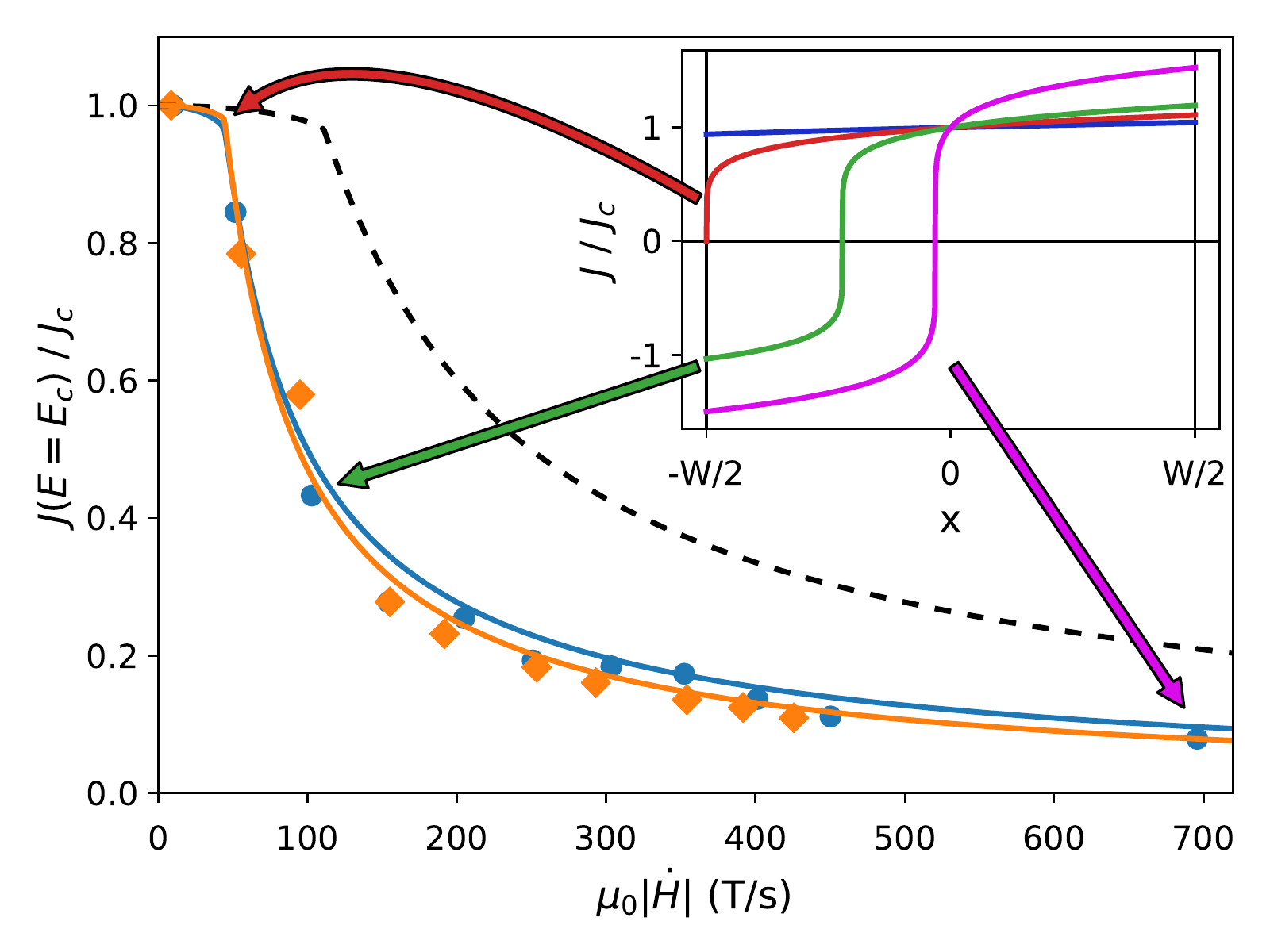}
\caption{\label{fig:JcBdot} \textbf{Dependence of the ``critical current'' on the magnetic field rate of change.} 
From the dependence of the I-V curves on $\dot{H}$ in Fig.~\ref{fig:IVBdot}, we report the current density at $E=E_c=11$\,\muV/cm as a function of $|\dot{H}|$.
\textbf{(orange diamonds)} YGBCO at 50\,K/10\,T. \textbf{(blue circles)} YGBCO+3BHO at 26.6\,K/30\,T.
\textbf{(solid lines)} Theoretical curves expected from Eq.\ref{Icalc}, which again show the excellent quantitative agreement with the experimental points.
\textbf{(black dashed line)} Theoretical curve for YGBCO+3BHO at 26.6K/30 T for a sample width $W =20$\,\mum, showing that narrower bridges should extend the range of $\dot{H}$ where $J(E=E_c)$ matches $J_c$ .
\textbf{(inset)} Theoretical current density profiles in YGBCO+3BHO at 26.6\,K/30\,T for $-\dot{H} = 15,\ 44,\ 100,\ 700$\,T/s. The drop of $J(E=E_c)$, as $|\dot{H}|$ crosses over $2E_c/W\approx 44$\,T/s, corresponds to the onset of current reversal at the edge of the sample (red curve).
 }
\end{figure}

The dramatic influence of $\dot{H}$ on the determination of \Jc\ can also be represented by reporting the current density at $E=E_c$ as a function of $\dot{H}$, as in Fig.~\ref{fig:JcBdot}.
For low $\dot{H}$ (i.e. $\epsilon < E_c$, $\mu_0|\dot{H}|<44$\,T/s) $J (E=E_c)$ stays constant and similar to $J_c(\dot{H}=0)$, but for higher $\dot{H}$ ($\epsilon > E_c$), $J (E=E_c)$ drops rapidly as the initial linear I-V component is present.
The effective $J_c$ curves derived from Eq.~\ref{Icalc} are displayed in Fig.~\ref{fig:JcBdot} and show an excellent agreement with the experimental data for both samples in different field regimes and temperature (50\,K - 10\,T  and 26.6\,K - 30\,T for YGBCO and YGBCO+3BHO samples respectively). The slight difference between the data sets comes from different $n$-values.
To illustrate the drastic change in behavior around $\epsilon = E_c $, we calculate the profile of the current density at $E=E_c$, across the width of the sample for different values of $\dot{H}$ (inset of Fig.~\ref{fig:JcBdot}). This highlights that the crossover at $\epsilon = E_c $ in the model, corresponds to the onset of a reversal in the direction of the current on one edge of the sample. The change in current distribution marks the crossover from linear I-V (with a slope that depends on $\dot{H}$) to a single, $\dot{H}$ independent, power-law behavior as can be clearly observed in the inset of Fig. \ref{fig:IVBdot}. This is important experimentally as it demonstrates that we can extrapolate back from the high $E$-high $J$ region of the I-V curve to obtain $J_c$.
 
\section{Discussion}

The dependence of the I-V shape with $dH/dt$ has a clear practical impact on the ability to determine $J_c(H)$ using the entire field pulse. The value of $W$ controls the shape of the I-V and the range of $\dot{H}$ where the I-V curves exhibit the standard power-law behavior. Specifically, the maximum $\dot{H}$ that can be used in determining $J_c$ (from $J$ at a given $E_c$) is $\mu_0\dot{H}_m = 2 \frac{E_c}{W}$, hence the narrower the width $W$, the larger the range of $\dot{H}$. 
For instance, the dashed black line in Fig.~\ref{fig:JcBdot} is the theoretical curve for the YGBCO+3BHO sample using the same parameters but a narrower bridge of width $W=20\,$\mum. The $\dot{H}$ range where standard I-V curves can be used directly is extended up to $\mu_0\dot{H}\approx 100$\,T/s. 
The range of standard I-V curves could also be extended by using magnetic field pulses with a flat-top (low $\dot{H}$ at peak field), which, for instance, can be achieved by using feedback controllers in a compact magnet or with the 60\,T long-pulse at the NHMFL.\cite{kohama60T,nhmfl}

Even if only the linear region of the I-V curves is available, these I-V curves can be fitted using Eq.~\ref{Icalc} with $J_c$ and $n$ as free parameters.  Our model enables us to use I-V curves that show an extended linear region to extract $J_c$ and the $n$-value, expanding the region of the magnetic field pulse where $J_c$ can be determined. We validate this idea by fitting the high $\dot{H}$ data of I-V displayed in Fig. \ref{fig:JcH} and obtaining $J_c$ and $n$ values that agree within 15 \% of those obtained for fitting the non-linear region. Because of the $(n-1)/n$ term in Eq. \ref{Reff}, this determination of $J_c$ works best for $n\gg 1$ when the fit is less sensitive to $n$ and highly stable and precise for $J_c$. Naturally, the higher the values of V used, the better the fit we obtain.  In summary, the understanding and modeling of the I-V curves in the high $dH/dt$ regime enables choosing the geometry to minimize its effects, measuring at high $E$-high $J$ and extrapolating back to $E_c-J_c$ values, or fitting the complete I-V dataset (as in Fig.\ref{fig:IVBdot}) even only in the linear regime.

Previous theoretical studies\cite{Brandt1993,Zeldov1994} considered the general case of a thin superconducting strip in a time-dependent perpendicular magnetic field with a time-dependent current. These studies noticed the importance of the asymmetry of the Bean profile in a sinusoidal AC magnetic field, which leads to AC losses via ``flux pumping'' across the sample. Although the pulsed-field experiment does not have ``flux pumping'' per se, it shares the physics that stems from the asymmetric Bean profile.
The specific case in our measurements could in principle be derived from these general cases, but this derivation, and especially the expression for the I-V curves, had not been developed to our knowledge.  

Remarkably, the I-V curves of Risse et al.\cite{Risse1997} in AC losses measurements resemble the I-V curves shown in Fig.\ref{fig:IVBdot}, even though the regime explored in Ref.~\cite{Risse1997} (low-field and weak-pinning) is very different from our high-field, strong-pinning regime.
In an attempt to explain Risse et al. results, Mikitik \& Brandt\cite{Mikitik2002} considered the case when the current is kept constant and an AC field is applied on top of a DC field, which shares similarities with our case. 
Because the focus of Ref.~\cite{Mikitik2002} is solving the field profile and the dependencies on the amplitude of the AC magnetic field, no numerical results or explicit derivations were reported for the I-V curves.
AC losses measurements\cite{Uksusman2009} in strong pinning YBCO superconductors also display an ohmic behavior above a threshold in current.  Although the asymmetry in the Bean profile is different from the pulsed-field case, we find similar values of the effective resistance using the model we present here. 

One assumption in our model is that the $B(x)$ profile maintains its ``V'' shape as $H $ increases or decreases, meaning that the vortices have enough time to adjust. 
This critical state can self-organize on a time scale set by the diffusion time $\tau_0$.
Gurevich and K\"{u}pfer~\cite{gurevich1993} showed that $\tau_0 \propto S\cdot J_c$, where $S = d\ln M/d\ln t $ is the normalized flux creep rate, and from experimental creep studies they also found that $\tau_0 \propto 1/\dot{H}$. 
For YBCO, they reported values of $\tau_0$ ranging from 1 to $10^4$\,s for $10^{-6}\dot{H}<10^{-2}~$T/s.
Assuming we can extend their results and analysis to our pulsed experiments with larger time-dependent $\dot{H}$ ($\approx10^2 - 10^3$ T/s), we obtain $\tau_0$ ranging from 100 to 10\,\mus\ at peak field and $\tau_0=1$\,\mus\ for $\dot{H}\approx 10^4$\,T/s, the fastest $\dot{H}$ during the upsweep before peak field. 
It thus seems justified that the critical state is established at a timescale shorter than the duration of our pulses of current, and that the asymmetric ``V''-shaped field profile moves as a block, unlike the case of AC loss studies where the field profile is reversed each cycle.

\section{Conclusions}
We performed non linear I-V curves measurements in pulsed magnetic fields up to 50\,T in YGBCO based coated conductors, showing excellent reproducibility and agreement with measurements in the same samples performed under DC magnetic fields.
This technique enables routine $J_c$ measurements up to 30\,T in table-top pulsed field systems\cite{30Tpulse} and opens up the exploration of superconductors solid vortex phase up to 100\,T, thus unlocking the access to a region of lower temperatures/higher fields where thermal fluctuation influence should diminish and quantum and field-induced disorder should compete. The smart I-V curve technique can also be applied to a large variety of condensed matter systems. \cite{Gerber3DCDWYBCOhighB,Ramshaw,Shin2017} 

Our $J_c$ measurements at high fields and $T=27~$K temperature indicate a rapid decrease near the irreversibility line, similarly to previous results at higher temperature and lower field. This supports the possibility of scaling $J_c$ by $H/H_{irr}$ as previously suggested\cite{civalejltp}. The knowledge of $J_c$ values at high fields is crucial for the development of future superconducting magnets and inserts for high field applications\cite{tokamak} and research\cite{progress32Tinsert,32Tmaglosses}. 
We also found a striking effect of $\dot{H}$ on the I-V curves and presented a model that accurately reproduces this effect.
The understanding and full description of this phenomenon is fundamental to design future experiments in superconductors at high pulsed magnetic fields. 
The experimental and theoretical understanding of vortex physics at large rate of change of the magnetic field has also a clear practical application for superconducting magnets technology and quench behavior.

\section{Acknowledgments}
We thank A. Gurevich and A. Koshelev for fruitful exchanges. Experiment design (B.M.), measurements (M.L., B.M.) and data analysis (M.L., L.C. and B.M.), are funded by the US DOE, Office of Basic Energy Sciences, Materials Sciences and Engineering Division. Work by F.F.B. (FPGA design, measurements) performed at NHMFL pulsed field facility at Los Alamos National Laboratory is funded by NSF by Grant No. 1157490/1644779. Work by K.A. and M.M. at Seikei University (sample fabrication) is supported by JSPS KAKENHI (17H 03239 and 17K 18888) and a research grant from the Japan Power Academy. dc measurements were performed at the Center for Integrated Nanotechnologies, an Office of Science User Facility operated for the U.S. DOE Office of Science. Measurements in pulsed fields were performed at the National High Magnetic Field Laboratory, which is supported by the National Science Foundation Cooperative Agreement No. DMR-1157490, the State of Florida and the United States Department of Energy.
 
\section{Methods}
{
\small
\subsection{Samples}
The samples of  \YGBCO\ (YGBCO) and \YGBCO\ with 3\% by weight of \BHO\ nanoparticles (YGBCO+3BHO) were grown using MOD on CeO$_2$ capped IBAD-MgO metal substrates, following Miura \textit{et al.}~\cite{Miura2017NatAsia}. The nominal $J_c$ in self-field at 77\,K on short samples were 5.3 and 5.0\,MA/cm$^2$ respectively.
To increase the total length of the bridge and reduce the open loop areas, we used laser lithography to etch a $50$\,\mum\ wide meandering pattern (see inset of Fig.~\ref{fig:raw_pulses}). Increasing the total length improves the voltage resolution to values close to the standard 1\,\muV/cm criterion. Reducing the open loop areas suppresses the voltage induced by the rapidly changing magnetic field, enabling better detection of the signal. 
Using a compensation loop the pulsed-field-induced voltage was thus reduced to the same order of magnitude as the signal. 
  
After laser lithography we observed a decrease in \Jc\ which is likely caused by small defects reducing the effective cross-section in the 4.51\,cm long and 50\,\mum\ wide meanders. We find self-field $J_c$ values of 3.3 and 1.5 MA/cm$^2$ at 77\,K for the YGBCO and YGBCO+3BHO samples respectively, however at high field and low temperature the \Jc\ values of the YGBCO+3BHO sample are typically twice as large as those of the YGBCO sample.
 
\subsection{FPGA-based I-V curves measurements in pulsed magnetic fields}

\setcounter{figure}{0}
\renewcommand{\thefigure}{M.\arabic{figure}}

\begin{figure}
\includegraphics[width=0.65\columnwidth]{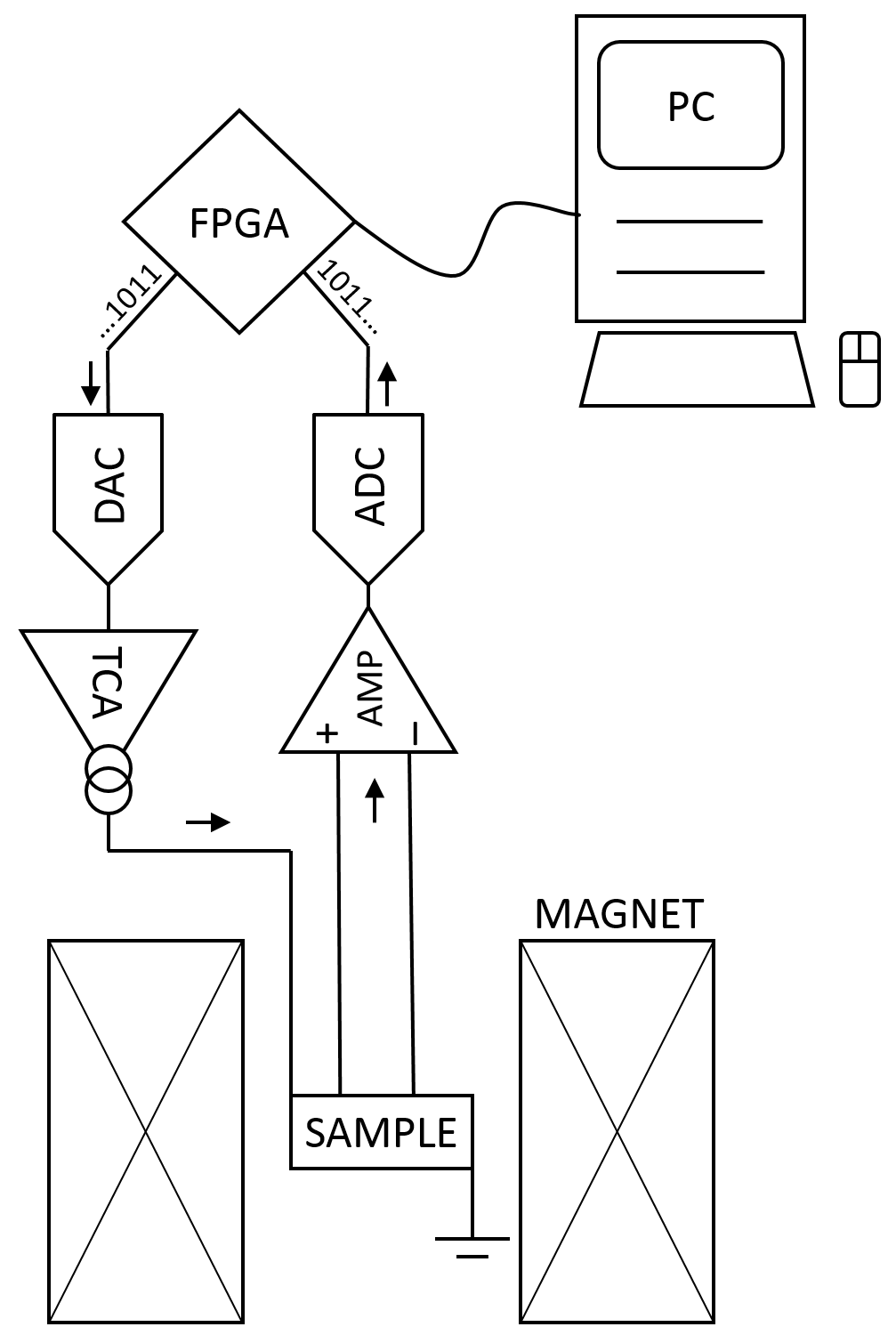}%
\caption{\label{fig:schem}\textbf{Schematic of the FPGA-based experimental setup} A computer readies the FPGA to output a series of user-defined voltage pulses. At the trigger of the pulsed magnet, the FPGA fires this series of pulses. These are converted to pulses of current via a high-speed trans-conductance amplifier (TCA) before passing into the superconducting sample. The voltage across the superconductor is simultaneously monitored by the FPGA via a low-noise preamplifier. If a set voltage threshold is exceeded at any pulses of current, the FPGA aborts or resets the current series of voltage pulses thus protecting the sample.
}
\end{figure}

A low-noise FPGA-based system enables taking multiple I-V curves in one field pulse while ensuring the voltage never exceeds a pre-determined maximum value to preserve the integrity of the sample\cite{Moll}.
The critical current detector is based on the Red Pitaya platform.\cite{RedPitaya} The detector system architecture is derived from Red Pitaya Notes open source code made available by Pavel Demin\cite{Demin}.
The custom user interface is implemented using National Instruments LabVIEW.
The Red Pitaya is programmed to produce a user-defined series of voltage pulses with increasingly larger amplitudes at 125\,MSPS.
These are converted to current pulses by a transconductance power amplifier (Valhalla Scientific) and applied to the sample.
The current pulse's length, shape, and amplitude pattern can be modified according to the measurement/sample requirements. 
To mitigate the inductive response of the sample strip which creates positive (negative) voltage spikes at the front (end) of each pulse, we used a half-sinewave shape for the pulse rise/fall (which reduces high-frequency content and thus the spike magnitude).
 
Simultaneously, the Red Pitaya records the voltage across the superconductor via a custom preamplifier based on Analog Devices' AD8429 with 1\,nV$/\sqrt{\mathrm{Hz}}$ noise figure.
A running sum comparator with background subtraction is implemented in the FPGA programmable logic.
The comparator measures the average voltage across the superconductor in response to each current pulse, and compares it against a set voltage threshold. 
The logic can detect if the voltage threshold is exceeded after each current pulse, thus protecting the sample against thermal runaway and eventual destruction.
Sample current is monitored through a 1.8 $\Omega$ resistor and recorded synchronously on another input channel of the Red Pitaya at a rate of 125\,MSPS.
Complete waveforms containing the sample voltage and current are recorded into the Red Pitaya memory during the field pulse. The waveform data is then uploaded into a client PC and re-analyzed with each current pulse averaging and background subtraction procedures re-processed, as shown in Fig.~\ref{fig:raw_pulses}. In this way, hundreds of I-V data points can be acquired in a single magnetic field pulse.

As in previous studies,\cite{miura} we rule out heating at the substrate or from rapid vortex movements as no variation with maximum field was found in AC-transport measurements in the normal and liquid vortex states with values of the magneto-resistance or irreversibility line unchanged.

 \subsection{Flux flow velocity and minimum crossing time}
From the flux-flow velocity, we find an upper bound for vortex velocity of $v_{ff}=8$\,km/s (Eq. 2.27 in Blatter\textit{ et al.}~\cite{BlatterRMP} with $\rho_n=50\times 10^{-6}\,\Omega$.cm, $J=0.5\times 10^6$\,A/cm$^2$ and $H_{c2}=9\times10^5\,$Oe).
This estimate of $v_{ff}$ is in agreement with vortex velocities of km/s observed in clean thin films of lead and in other clean materials (see Embon \textit{et al.}~\cite{embon2017} and references therein).
This velocity sets to 6\,ns the minimum, physically possible, characteristic time for vortices to cross a $50$\,\mum-wide bridge of YGBCO+3BHO.

\subsection{CGS units}
Here we present the equivalent version in cgs units of the main text equations in SI units.

\noindent
Vortex velocity produced by $\dot{H}$:
$$
\vec{v}_{\dot{H}}(x)  = - \frac{\dot{H}}{B(x)} \, x \, \vec{e_x}
$$
Electric field produced by $\dot{H}$:
$$
\vec{E}_{\dot{H}}(x) =\frac{1}{c} \vec{B} \times \vec{v} = - \frac{1}{c} \dot{H} \, x \, \vec{e_y}
$$
Current density produced by $\dot{H}$:
$$
\vec{J}_{\dot{H}}(x) = J_c \cdot \left| \frac{x \dot{H} }{c E_c }\right|^{1/n} \cdot sign(- \dot{H} x)\, \vec{e_y}
$$
Amp\`ere-Maxwell:
$$
\frac{\partial B_z}{\partial x} = - \frac{4\pi}{c} J_y
$$ 
Magnetic field profile produced by $\dot{H}$:
$$
B(x) = H + \frac{4\pi}{c} J_c\left(\frac{\dot{H}}{cE_c}\right)^{1/n} \frac{n}{n+1} \left[\left(\frac{W}{2}\right)^{\frac{n+1}{n}}-\left|x\right|^{\frac{n+1}{n}}\right].
$$
Total external DC current I passing through the sample in the presence of $\dot{H}$:
$$
I = \frac{d W}{2\epsilon} \frac{J_c}{E_c^{1/n}} \frac{n}{n+1} \left[\left|E_J+\epsilon E_c \right|^{\frac{n+1}{n}}-\left|E_J-\epsilon E_c \right|^{\frac{n+1}{n}}\right]
$$
where $\epsilon = \frac{\dot{H} W}{2 c E_c} $ and $E_J = \frac{V}{L}$ is the measured electrical voltage.
}
 
\renewcommand{\thefigure}{M.\arabic{figure}}
\begin{figure}
\includegraphics[width=\columnwidth]{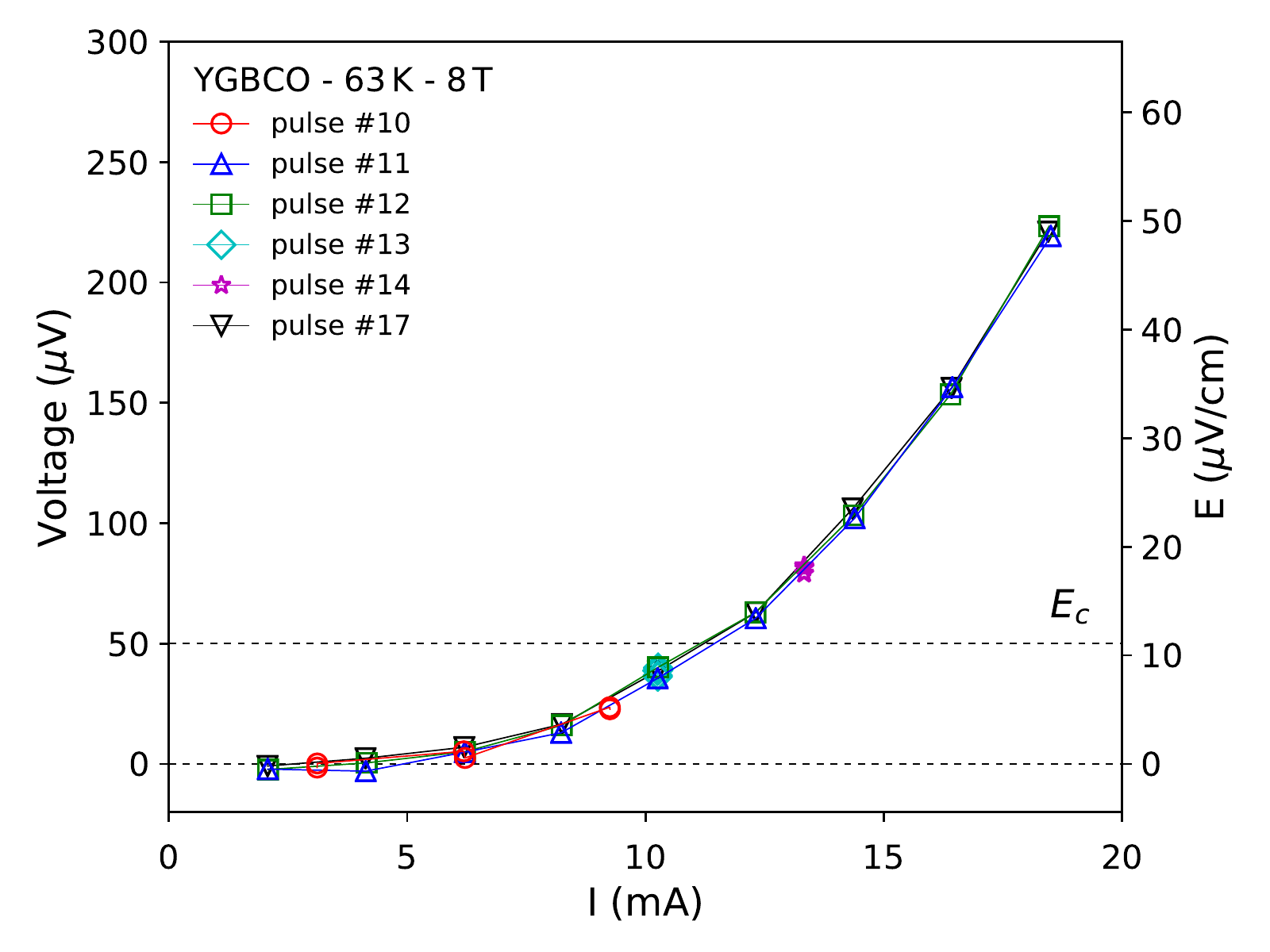}%
\caption{\label{fig:reproducible}\textbf{Reproducible and robust current-voltage curves.} I-V curves measured in a YGBCO sample at 63\,K and 8\,T in pulsed magnetic fields. 
Each curve corresponds to data taken at peak field during a single magnetic field pulse.
This data shows that the I-V curves are reproducible and robust against variations of the acquisition parameters: \#10: 3\,mA steps (i.e. 0.03 MA/cm$^2$) and 19.2\,\mus\ voltage averaging on-board the FPGA; \#11: 2\,mA steps and 19.2\,\mus\ integration; \#12 same as \#11; \#13: 10\,mA steps and 19.2\,\mus\ integration; \#14: 13\,mA steps and 19.2\,\mus\ integration; \#17: 2\,mA steps and 12.8\,\mus\ integration.
}
\end{figure}

\section*{References}
\vspace{-5mm}
\bibliography{biblio}

\end{document}